# Coupling Quantum Emitters in WSe$_2$ Monolayers to a Metal-Insulator-Metal Waveguide

*Subhojit Dutta,[a] Tao Cai,[a] Mustafa Atabey Buyukkaya,[a] Sabyasachi Barik,[a,b] Shahriar Aghaeimeibodi,[a] Edo Waks[a,b]*

[a] Department of Electrical and Computer Engineering and Institute for Research in Electronics and Applied Physics, University of Maryland, College Park, Maryland 20742, USA.

[b] Joint Quantum Institute, University of Maryland and the National Institute of Standards and Technology, College Park, Maryland 20742, USA.

\* Corresponding author E-mail: edowaks@umd.edu

**ABSTRACT:** Coupling single photon emitters to surface plasmons provides a versatile ground for on chip quantum photonics. However, achieving good coupling efficiency requires precise alignment of both the position and dipole orientation of the emitter relative to the plasmonic mode. We demonstrate coupling of single emitters in the 2-D semiconductor, WSe$_2$ self-aligned with propagating surface plasmon polaritons in silver-air-silver, metal-insulator-metal waveguides. The waveguide produces strain induced defects in the monolayer which are close to the surface plasmon mode with favorable dipole orientations for optimal coupling. We measure an average enhancement in the rate of spontaneous emission by a factor of 1.89 for coupling the single defects to the plasmonic waveguide. This architecture provides an efficient way of coupling single photon emitters to propagating plasmons which is an important step towards realizing active plasmonic circuits on chip.

In recent years, defects bound excitons in two dimensional semiconductors have emerged as a new class of single photon emitters with ultra-narrow linewidths of 100 μeVs, as well as high single photon purity.[1–4] These emitters are located at the surface of an atomically thin monolayer which allows them to come in close proximity to photonic nanostructures. Another characteristic feature is that they can be deterministically induced by strain engineering allowing for site specific positioning.[5–7] Thus, with the ability to position these atomically thin quantum emitters, one can efficiently couple to the confined mode of optical nanostructures providing a platform for coherent light-matter interactions. Such a platform is critical for applications such as quantum communication and quantum information processing[8,9].

2D materials with their narrow linewidths and flexibility in terms of positioning, offer a promising path to tailoring strong light matter interactions. However, apart from nanoscale positioning of emitters, the nanostructure must also exhibit a high optical density of states. A strong contender for the photonic nanostructures are surface plasmon polaritons generated at a metal-dielectric interface. Surface plasmons exhibit extreme subwavelength confinement of light[10,11] and an atom-like dipole emitter placed near the metal-dielectric interface, preferentially emits into the surface plasmon mode due to its high optical density of states.[12] The strong optical decay of emitters into the surface plasmon results in efficient coupling of emitters to a common plasmonic mode that can lead to strong photon-photon interactions.[13] Coupling also produces a significant enhancement in the rate of spontaneous emission of the emitters[12,14,15] which can help realize a fast single photon source on-chip. Thus, single photon emitters in 2D materials, coupled to surface plasmon polaritons establishes a platform for compact active photonic circuits essential for quantum information processing[16–18].

Several previous works reported deterministic coupling of quantum emitters with plasmonic nanocavities[19–21]. However, this does not result in a propagating surface plasmon polariton which is desirable for non-linear plasmonic circuit elements. More recent work realizes a travelling mode by coupling single localized defects in a $WSe_2$ monolayer self-aligned to the propagating surface plasmon mode of a silver nanowire.[22] However, these emitters showed no significant radiative enhancement, because the electric field vector is perpendicular to the metal surface and directed radially outward from the cylindrical metal nanowires, whereas single emitters in 2D materials have in plane dipole moments [1–4] which are tangential to the metal surface. Thus, dipole emitters in a monolayer placed on top of a nanowire waveguide have an unfavorable alignment to the local electric field vector. Metal-Insulator-Metal (MIM) waveguides solve this problem because the electric field of the surface plasmon has a strong in-plane component that is perpendicular to the metal surface.

Here we show coupling of single emitters in $WSe_2$ with propagating surface plasmon polaritons in silver-air-silver, MIM waveguides. The strain gradient enforced on the monolayer by the waveguide generates sharp localized single photon emitters close to the plasmonic mode. The component of the in-plane dipole moment perpendicular to the metal surface preferentially couples to the travelling waveguide mode, leading to an average radiative lifetime enhancement of 1.89. A single emitter in a 2D material coupled to an MIM waveguides can lead to deterministic active plasmonic circuits that can be lithographically fabricated on chip.

Fig. 1(a) shows a finite difference eigenmode simulation for the surface plasmon mode in a silver-air-silver MIM structure. The simulation is performed using the commercial software, Lumerical MODE Solutions. We refer to the Palik model[23] for the silver films used in our simulations of the waveguide mode. The color map represents the magnitude of the electric field ($E_x$) vector of the plasmon polariton mode, directed along the X axis which is oriented in the in-plane direction with respect to the monolayer. Fig. 1(c) plots a vector map of the electric field showing a strong localization along the X axis in the air gap region between the two metal strips of the MIM waveguide. Because the monolayer dipole moment is in-plane, it can align to the electric field orientation of the waveguide. This contrasts with nanowire waveguides where the preferred dipole orientation is radial to the structure and always orthogonal to the plane of the monolayer, hampering coupling efficiency.

We fabricate the MIM waveguides using electron beam lithography, followed by metal deposition and liftoff. We spin coat $Si/SiO_2$ sample with ZEP520A ebeam resist at 4500 rpm and post bake it for 5 min at 180° C. Next, we pattern the samples using an electron beam lithography system at an acceleration voltage of 100 kV (you can cite the exact ebeam model here in parenthesis if you want), using a dose array ranging from 250 to 480 µC/cm$^2$. We develop the resist in n-Amyl Acetate (ZED-N50), Methyl isobutyl ketone and Isopropyl alcohol for 1min, 30s and 30s respectively. For the metal evaporation step, we use a thermal evaporator to deposit 5nm and 65nm, Cr and Ag at evaporation rates of 5 A/s and 130 A/s, respectively to achieve high quality plasmonic films.[24] We perform liftoff by soaking in acetone overnight and subsequently rinsing with Acetone and Isopropyl alcohol . The fabricated waveguides have gaps ranging from 90 nm to 110 nm and a length of 7 µm, Fig. 2(a). We cover the waveguides with a 4nm buffer layer of oxide to protect them from damage as well as to prevent significant effects of quenching.[25] After waveguide fabrication, we transfer the monolayer samples onto the MIM structures. We grow WSe$_2$ monolayers on a sapphire substrate using a chemical vapor deposition method[26] and then dry stamp them on the MIM sample using a polydimethylsiloxane (PDMS) gel as an intermediate transfer medium.[27]

To characterize the sample, we cool it to 3.2 K in closed cycle cryostat (Attocube Inc.) and perform photoluminescence measurements using a confocal microscope geometry. We use a continuous wave excitation laser emitting at 532 nm to perform photoluminescence spectroscopy. We focus the laser onto the sample surface with an objective lens (Numerical aperture 0.7) to excite a small diffraction limited spot on the MIM waveguide. A 700 nm long-pass optical filter rejects the pump wavelength to isolate the fluorescence signal. A monochrome scientific camera (Rolera-XR, Qimaging, Inc.) images the fluorescence field pattern. Alternately, we use a grating spectrometer (SP2750, Princeton Instruments) to measure the fluorescence spectrum.

Fig. 2(b) shows the photoluminescence map of the flake/waveguide system as observed on the camera. We observe bright localized emission spots in the vicinity of the waveguide. The photoluminescence spectrum, Fig. 2(c) of the bright spot as marked in Fig. 2(b) (dotted circle) shows the presence of several sharp peaks corresponding to single defects emitters in WSe$_2$.[1–4] These emitters arise from the strain gradient affected on the monolayer by the MIM waveguide. Such strain-induced defect formations have been previously shown in monolayers suspended over holes or placed on top of nanopillars and nanowires.[5,7,22] We further observe bright spots at the ends of the waveguide. We assert that the emission from these single defects couple to the propagating surface plasmon polariton in the waveguide and scatter of the ends as photons resulting in such bright spots.

To verify the assertion of coupling, we search for defect emissions at one end of the waveguide while looking for the scattered emission at the other end. Fig. 2(d) shows a representative camera image of the fluorescence intensity when we excite point A at one end of a waveguide (Waveguide 2). Fig 3(a) shows the photoluminescence spectrum with both the excitation and collection spots focused at A. We observe two resolvable peaks at 736.6 nm and 737.1 nm which are labelled as Z and X, respectively. We fit the peaks to a Lorentzian to determine linewidths of 0.17 nm (Z) and 0.10 nm (X). These peaks constitute a set of orthogonally polarized doublets.[1–4] (Supplementary material Fig. S3) Keeping the excitation at the location of the defect (end A) we move the collection spot to the far end of the waveguide (end B) and collect the corresponding spectrum, Fig 3(d). This time we observe only one of the doublet peaks, X at 737.4 nm since only one of the two cross polarized peaks aligns with the direction of the electric field of the surface plasmon polariton. The peak vanishes when we move the excitation spot away from A. The signal of the scattered emission (Fig. 3(d)) is weaker than the signal at the location of the defect (Fig. 3(a)). Thus, we integrate over a much longer period to obtain a high signal to noise ratio of 35.24. We observe a small redshift of 0.3 nm in the wavelength of the waveguided peak, X. To further understand this phenomenon, we measure similar spectra for three other coupled defects, P, Q and R on two different waveguides, Waveguide 3 and Waveguide 4 respectively. Fig 3(b) and (c) show the spectra of defects P, Q and R when both the excitation and collection are aligned to the location of the defects at one end of waveguides 3 and 4 respectively. Fig 3(d) and (e) show the spectra of the scattered defects when we move the collection point to the far end of the waveguides keeping the excitation at the same point. We observe a 0.11 nm redshift for defect P and a 0.15 nm blueshift for both defects Q and R when looking at the scattered spectra. We conclude that these small and consistent deviations are due to spectral wandering of the defect peaks in time over the course of the measurement. Indeed previous literature reports spectral wandering of upto 1 meV for such defects in $WSe_2$.[1–4] To further ensure that the emission we see is not due to the pump coupling to the waveguide and exciting a different defect at the other end, we move the excitation away from the location (A) of the defect and directly excite B in waveguide 2, measuring the emission at B. The spectrum shows the background photoluminescence emission from the monolayer flake and does not show the sharp quantum dot like feature. (Detailed discussion in supplementary material Fig. S4)

Spontaneous emission enhancement is a strong signature of efficient coupling of dipole emitters to optical nanostructures. But measuring radiative enhancement in $WSe_2$ emitters is complicated by the fact that even in pristine monolayer, the radiative lifetimes are broadly distributed and can range from 3 ns to as long as 19 ns.[21] To explore the enhancement in the rate of spontaneous emission, we therefore accumulate statistics over a total of 23 emitters, with 10 emitters located on the waveguide/monolayer system and 13 emitters located on the pristine monolayer, away from the waveguide. The lifetime measurements are carried out with the excitation and collection spots aligned at the same point, the location of the defect. The distributions of the radiative lifetimes for the two sets of emitters are plotted in Fig. 4(b) and (d). We see a clear decrease of the lifetimes for the emitters that are located on the MIM waveguide. The radiative lifetimes of the emitters are fitted to a Gaussian. We find that the emitters on the waveguide have an average lifetime of 3.18 ± 1.12 ns while those off the waveguide have an average lifetime of 6.01 ± 0.94 ns, the error bars representing the 95% confidence bounds for the means.. Thus, we can identify two distributions with distinctly resolvable means. From these values we determine an average radiative lifetime enhancement of 1.89 for near-field coupling of the single emitters to the MIM waveguides.

In conclusion, we showed coupling of single emitters in $WSe_2$ with propagating surface plasmon polaritons in MIM waveguides. The strain gradient enforced on the monolayer by the waveguide generated localized single photon emitters close to the plasmonic mode. The coupled emitters experienced an enhancement in the average radiative lifetime measured by a factor of 1.89. Our results can realize a fast, integrated single photon source by coupling the MIM structure with a dielectric waveguide.[28,29] It is however, necessary to gain better control over the coupling strength between the emitter and the plasmonic mode. One approach might be to engineer better quality of defects by reducing the non-radiative relaxation in the system by encapsulating the monolayer by boron nitride layers[30] which is known to decrease the linewidths of the defects. With better control over the coupling strength one can think of more appealing applications that harp on the giant non-linearity presented by the plasmon-emitter system. The saturable two-level system of the emitter absorbs and subsequently scatters off single photons while being invisible to the next photon. This can realize a single photon switch.[18] Thus, combined with recent advances in the fabrication quality of 2D monolayer,[30] our results present a versatile platform which can lead to compact active plasmonic circuits on chip.

**Supplementary Material:** See supplementary material for further physical characterization of the waveguide by means of Scanning Electron Microscopy and Atomic Force Microscopy, for data regarding the cross polarized peaks X and Z as well as for a detailed discussion ruling out the issue of pump coupling.

**Acknowledgements:** The authors acknowledge support from the National Science Foundation (award number ECCS1508897), the Office of Naval Research ONR (award number N000141410612), the Air Force Office of Scientific Research (AFOSR) (award number 271470871D), and the Physics Frontier Center at the Joint Quantum Institute.

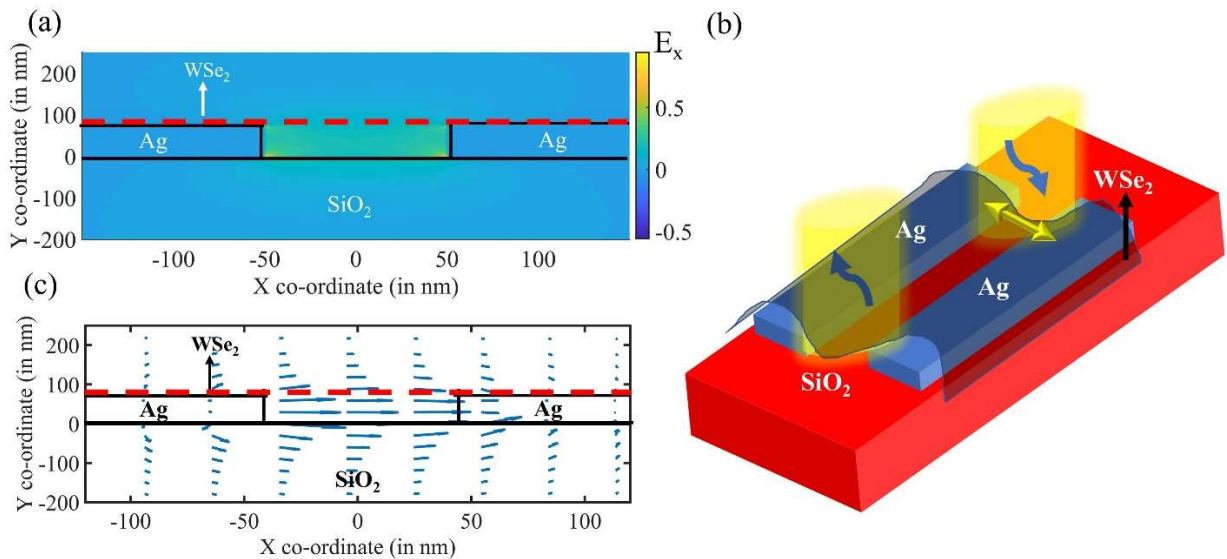

Fig. 1. (a) Finite Difference Eigenmode simulation (Lumerical Inc.) showing the X component of the electric field for the surface plasmon mode propagating along the Z axis in the MIM waveguide. (b) Schematic of an MIM waveguide covered by a $WSe_2$ monolayer. The yellow dipole represents a quantum emitter in $WSe_2$. The blue arrows denote the excitation and collection points

respectively. (c) Finite Difference Eigenmode simulation (Lumerical Inc.) showing the vector plot of the electric field in the gap region of the MIM waveguide.

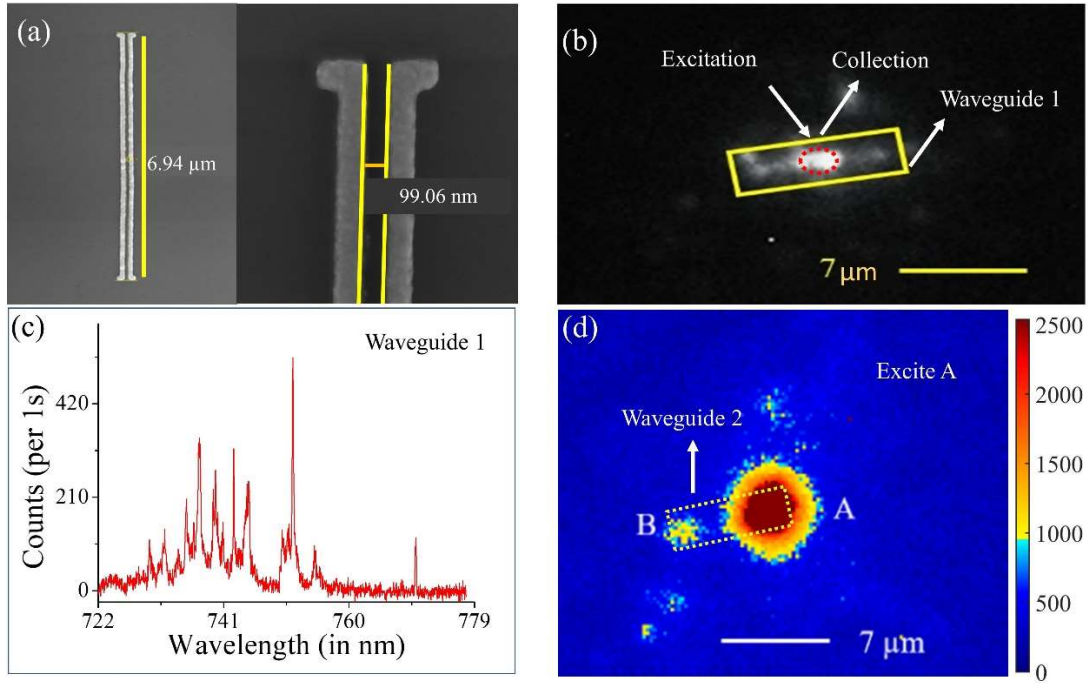

Fig. 2. (a) SEM image of a representative MIM waveguide showing the length and gap width. (b) Shows a photoluminescence intensity map of a $WSe_2$ flake located on top of a waveguide marked by the yellow box. (c) Photoluminescence spectrum collected at the point on the waveguide marked by the dotted circle in (b), showing multiple defect emissions structurally aligned with the waveguide. (d) Photoluminescence intensity map showing a representative flake/waveguide system while exciting a coupled defect located at one end A of a waveguide. The defect at A couples to the waveguide and is scattered off at B causing the bright feature.

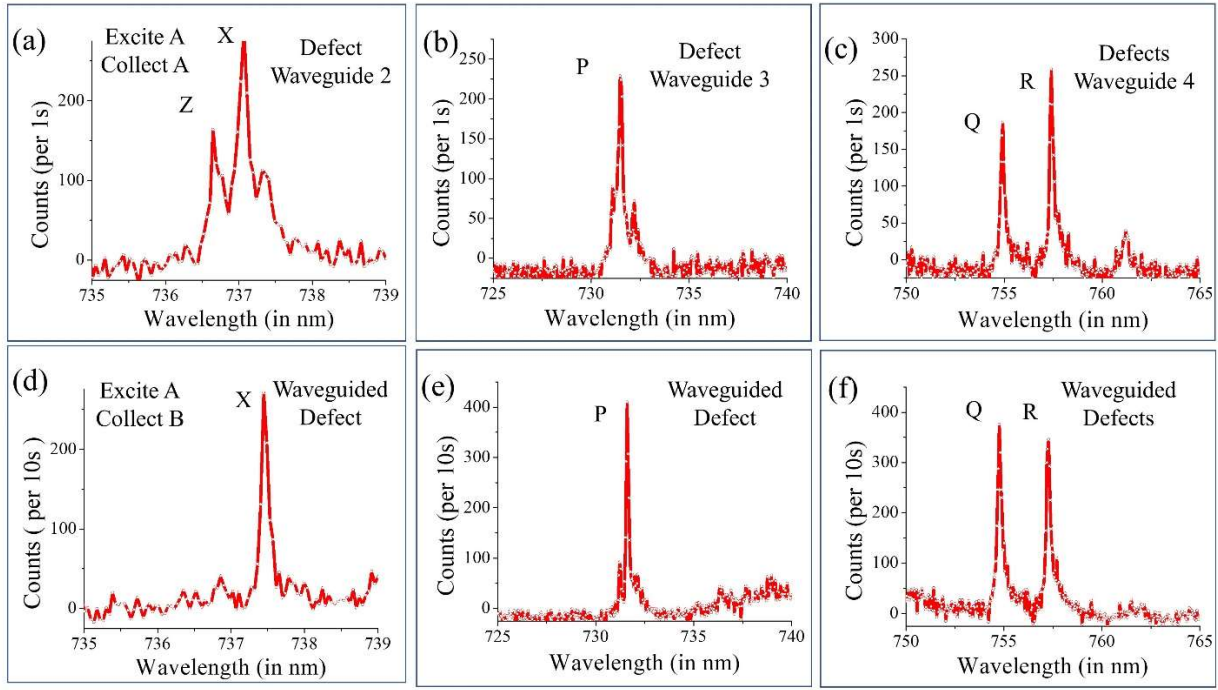

Fig. 3. (a), (b), (c) Shows the photoluminescence spectra of four different coupled defects (X, P, Q and R) with the excitation and collection spots aligned to the same point, at one end of three different waveguides. (d), (e), (f) Shows the photoluminescence spectra of the waveguided defects (X, P, Q and R) with the excitation spot fixed at the location of the defect and collection spot moved to the far end of the waveguides.

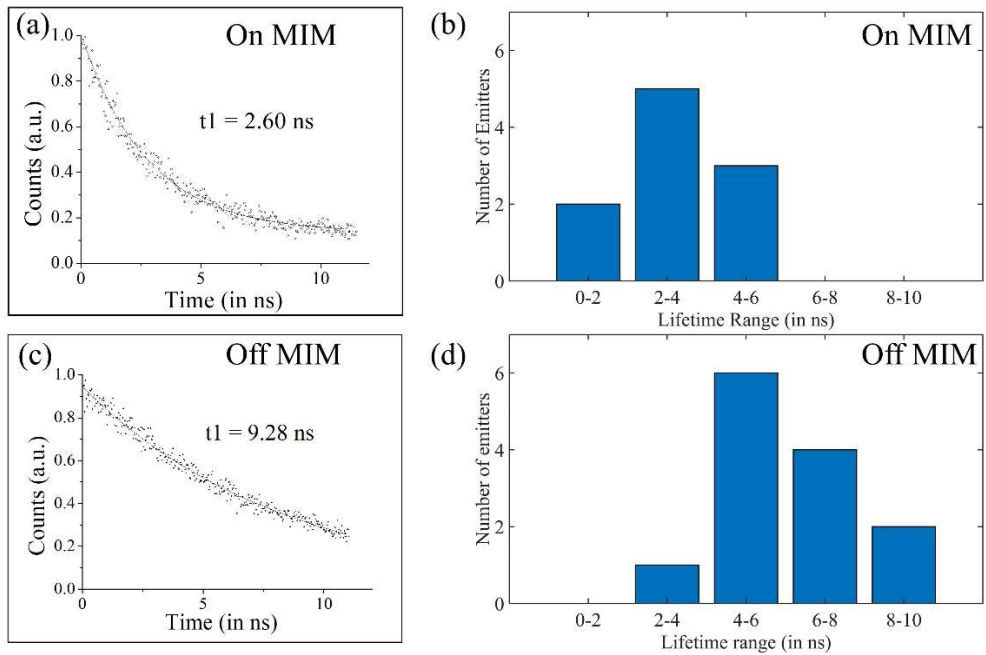

Fig. 4. (a) Lifetime of a representative emitter located on the MIM waveguide. (b) Lifetime statistics of emitters located on the waveguide. (c) Lifetime of a representative emitter located away from the MIM waveguide. (d) Lifetime statistics of emitters located far from the waveguide.